\documentstyle[12pt,aaspp4,flushrt]{article}

\newcommand{\apg}{\gtrsim}
\newcommand{\apl}{\lesssim}
\newcommand{\cmjj}{\mbox{${\rm cm^{-2}}$}}
\newcommand{\etal}{et al.}
\newcommand{\hI}{\mbox{${\rm H\ I}$}}

\newcommand{\ibid}{\underline{\makebox[0.5in]{}}.}

\newcommand{\kms}{\mbox{km\ s${^{-1}}$}}
\newcommand{\lya}{\mbox{${\rm Ly}\alpha$}}

\newcommand{\civ}{\mbox{${\rm C\ IV}$}}
 
\begin{document}
 
\lefthead{Chen et al.}
 
\righthead{}

 
\title{THE ORIGIN OF C IV ABSORPTION SYSTEMS AT REDSHIFTS $z<1$---DISCOVERY 
OF EXTENDED C IV ENVELOPES AROUND GALAXIES\altaffilmark{1}}
\altaffiltext{1}{Based on observations with the NASA/ESA Hubble Space
Telescope, obtained at the Space Telescope Science Institute, which is operated
by the Association of Universities for Research in Astronomy, Inc., under NASA
contract NAS5--26555.}
 
\author{HSIAO-WEN CHEN\altaffilmark{2} and KENNETH M. LANZETTA}
\affil{Department of Physics and Astronomy, State University of New York at
Stony Brook \\
Stony Brook, NY 11794--3800, U.S.A. \\
lanzetta@sbastr.ess.sunysb.edu}

\altaffiltext{2}{Current address: Observatories of the Carnegie Institution
of Washington, 813 Santa Barbara Street, Pasadena, CA 91101, U.S.A.  E-Mail: 
hchen@ociw.edu}

\and

\author{JOHN K. WEBB}
\affil{School of Physics, University of New South Wales \\
Sydney 2052, NSW, AUSTRALIA \\
jkw@edwin.phys.unsw.edu.au}

\newpage

\begin{abstract}

  We report the discovery of extended \civ\ gaseous envelopes around galaxies 
of a wide range of luminosity and morphological type.  First, we show that
\civ\ absorption systems are strongly clustered around galaxies on velocity
scales of $v \apl 250$ \kms\ and impact parameter scales of $\rho \apl 100 \
h^{-1}$ kpc but not on larger velocity or impact parameter scales.  Next,
adopting measurements of galaxy properties presented in previous papers (which
include $B$-band luminosity, surface brightness, and disk-to-bulge ratio),
we examine how properties of the \civ\ absorption systems depend on properties
of the galaxies.  On the basis of 14 galaxy and absorber pairs and 36 galaxies
that do not produce corresponding \civ\ absorption lines to within sensitive
upper limits, we find that:  (1) Galaxies of a range of morphological type and
luminosity appear to possess extended \civ\ gaseous envelopes of radius $R
\approx 100 \ h^{-1}$ kpc, with abrupt boundaries between the C\,IV absorbing
and non-absorbing regions.  (2) The extent of \civ-absorbing gas around
galaxies scales with galaxy $B$-band luminosity as $R \propto L_B^{0.5 \pm
0.1}$ but does not depend strongly on galaxy surface brightness, redshift,
or morphological type.  And (3) the covering factor of \civ\ clouds within
$\approx 100 \ h^{-1}$ kpc of galaxies is nearly unity, but there is a large
scatter in the mean number of clouds encountered along the line of sight.
After scaling to the luminosity of an $L_*$ galaxy, we find that 13 of 14
galaxies of impact parameter $\rho < 100 \ h^{-1}$ kpc are associated with
corresponding \civ\ absorption lines, while only one of 36 galaxies of impact
parameter $\rho > 100 \ h^{-1}$ kpc are associated with corresponding \civ\
absorption lines.  The most significant implication of the study is that
galaxies of a wide range of luminosity and morphological type are surrounded by
chemically enriched gas that extends for at least $\approx 100 \ h^{-1}$ kpc.
We consider various scenarios that may have produced metals at large galactic 
distance and conclude that accreting satellites are most likely to be 
responsible for chemically enriched gas at large galactic distances to regular
looking galaxies.

\end{abstract}

\keywords{galaxies: evolution---quasars: absorption lines}

\newpage

\section{INTRODUCTION}

  Metal-line absorption systems observed in the spectra of background QSOs 
provide a unique probe of the chemical content and dynamics of gas at large
galactic distances.  Comparison of galaxies and Mg\,II absorption systems along
common lines of sight has demonstrated that (1) Mg\,II absorption systems arise
in extended gaseous envelopes of galaxies and (2) the gaseous extent of
galaxies (at very low column densities) stretches for many times the optical
extent of galaxies (Bergeron \& Boiss\'e 1991; Lanzetta \& Bowen 1990, 1992;
Steidel 1993).  But while occasional galaxies associated with \civ\ absorption
systems have been reported by various authors (Bergeron \etal\ 1994; Lanzetta
\etal\ 1995; Steidel \etal\ 1997; Churchill \etal\ 1999), no uniform sample of
galaxies and \civ\ absorption systems has yet been applied to systematically
investigate statistical properties of extended \civ-absorbing gas around
galaxies.  In contrast to Mg\,II absorption systems, C\,IV absorption systems 
probe highly-ionized, low-density regions (Wolfe 1983; Bergeron \& Stasi\'nska
1986), which suggests that gas traced by C\,IV is typically at much larger
galactic distances than gas traced by Mg\,II.  Hence C\,IV absorption systems
may bear importantly on understanding the processes of galaxy formation and
evolution that deposit chemically enriched material far from galaxies.

  Over the past several years, we have been conducting an imaging and
spectroscopic survey of faint galaxies in fields of Hubble Space Telescope
(HST) spectroscopic target QSOs (Lanzetta \etal\ 1995; Chen \etal\ 1998, 2001, 
hereafter Papers I and II).  The goal of the survey is to determine the gaseous
extent of galaxies and the origin of QSO absorption systems by directly
comparing galaxies and QSO absorption systems along common lines of sight.  As
a result of the survey, we have so far identified 352 galaxies of apparent
magnitude $m_R < 23$ and redshift $z < 1.2$, 230 \lya\ absorption systems of
redshift $z < 1.1$, and 36 \civ\ absorption systems of redshift $z < 0.9$ in 24
QSO fields.  Impact parameters of the galaxies to the QSO lines of sight range
from $\rho = 10.9$ to $1576.7 \ h^{-1}$ kpc.  Our galaxy and absorber sample
provides for the first time the opportunity to study statistical properties of
\civ-absorbing gas around galaxies.

  Here we present results of the study.  First, we show that \civ\ absorption
systems are strongly clustered around galaxies on velocity scales of $v \apl
250$ \kms\ and impact parameter scales of $\rho \apl 100 \ h^{-1}$ kpc but not
on larger velocity or impact parameter scales.  This demonstrates that C\,IV
absorption systems are associated with galaxies and strongly suggests that
C\,IV-absorbing gas generally arises in individual galaxies, rather than in a
diffuse medium that is loosely associated with galaxy groups, clusters, or
other large-scale structures.  Next, adopting measurements of galaxy properties
presented in Papers I and II (which include $B$-band luminosity, surface
brightness, and disk-to-bulge ratio), we examine how properties of the \civ\
absorption systems depend on properties of the galaxies.  On the basis of 14
galaxy and absorber pairs and 36 galaxies that do not produce corresponding
\civ\ absorption lines to within sensitive upper limits, we find that:  (1)
Galaxies of a range of morphological type and luminosity appear to possess
extended \civ\ gaseous envelopes of radius $R \approx 100 \ h^{-1}$ kpc, with
abrupt boundaries between the C\,IV absorbing and non-absorbing regions.  (2)
The extent of \civ-absorbing gas around galaxies scales with galaxy $B$-band
luminosity as $R \propto L_B^{0.5 \pm 0.1}$ but does not depend strongly on
galaxy surface brightness, redshift, or morphological type.  And (3) the
covering factor of \civ\ clouds within $\approx 100 \ h^{-1}$ kpc of galaxies
is nearly unity, but there is a large scatter in the mean number of clouds
encountered along the line of sight.  After scaling to the luminosity of an
$L_*$ galaxy, we find that 13 of 14 galaxies of impact parameter $\rho < 100 \
h^{-1}$ kpc are associated with corresponding \civ\ absorption lines, while
only one of 36 galaxies of impact parameter $\rho > 100 \ h^{-1}$ kpc are
associated with corresponding \civ\ absorption lines.

  Our results indicate that galaxies of a wide range of luminosity and
morphological type are surrounded by chemically enriched gas that extends for
at least $\approx 100 \ h^{-1}$ kpc.  Given such a large extent, we consider it
unlikely that the absorbing clouds were ejected from the primary galaxies by
stellar winds or galactic fountains.  Rather, we conclude that the absorbing
clouds were probably produced at large galactic radii by either Population III
stars or by progressive accretion of gas from surrounding satellite galaxies.
If the clouds were formed by satellite accretion, then the abrupt boundaries
between the \civ\ absorbing and non-absorbing regions suggest that galaxies
form through dissipational accretion, and the lack of a strong redshift
dependence of the extent of \civ-absorbing gas around galaxies suggests that
the typical separation between satellites and primary galaxies is larger than
100 $h^{-1}$ kpc.  We adopt a standard Friedmann cosmology of dimensionless
Hubble constant $h = H_0/(100\ {\rm km} \ {\rm s}^{-1} \ {\rm Mpc}^{-1})$ and
deceleration parameter $q_0 = 0.5$ throughout.

\section{DATA}

   The analysis is based on observations of our ongoing imaging and 
spectroscopic survey of faint galaxies in fields of HST spectroscopic target 
QSOs (Lanzetta \etal\ 1995; Papers I and II; Lanzetta \etal\ 2001 in 
preparation).  As a result of the survey, we have so far identified 352 
galaxies of apparent magnitude $m_R < 23$ and redshift $z < 1.2$, 230 \lya\ 
absorption systems of redshift $z < 1.1$, and 36 \civ\ absorption systems of 
redshift $z < 0.9$ in 24 QSO fields.  Impact parameters of the galaxies to the 
QSO lines of sight range from $\rho = 10.9$ to $1576.7 \ h^{-1}$ kpc.

  Using the Hubble Space Telescope with the Wide Field and Planetary Camera 2,
we have also obtained high-quality optical-wavelength images of 142 of the
galaxies in 19 fields, which we have used to measure galaxy $B$-band
luminosity, effective radius, surface brightness, inclination and orientation 
of the disk component, axial ratio of the bulge component, and disk-to-bulge 
ratio by means of a two-dimensional surface brightness profile analysis.  These
measurements are described and presented in Papers I and II.

\section{ANALYSIS}

  In this section, we examine how properties of the \civ\ absorption systems
depend on properties of the galaxies.  
 
\subsection{The Galaxy--C IV Absorber Cross-Correlation Function}

  To establish the statistical relationship between galaxies and \civ\
absorption systems, we measured the galaxy--\civ\ absorber cross-correlation
function $\xi_{\rm ga}(v,\rho)$ as it depends on line-of-sight velocity
separation $v$ and impact parameter separation $\rho$.  Figure 1 shows
$\xi_{\rm ga}(v,\rho)$ determined from 352 galaxies and 36 \civ\ absorption
systems.  The bin size in velocity separation is 250 \kms, and the bin size in
impact parameter separation is indicated in each panel.  A total of 563 galaxy
and absorber pairs enter into the analysis, of which 153 are of impact
parameter separation $\rho < 100 \ h^{-1}$ kpc, 148 are of $ 100 < \rho < 200 \
h^{-1}$ kpc, and 262 are of $\rho > 200 \ h^{-1}$ kpc.  Error bars indicate 1
$\sigma$ Poisson counting fluctuations.  

  Figure 1 indicates a statistically significant excess of galaxy and \civ\
absorber pairs at velocity separations $v \apl 250$ \kms\ and impact parameter
separations $\rho \apl 100 \ h^{-1}$ kpc, but no excess at larger velocities
or impact parameters.  Specifically, the bin spanning $v < 250$ \kms\ and
$\rho < 100 \ h^{-1}$ kpc is expected to contain $0.82 \pm 1.34$ pairs but is
observed to contain 9 pairs, which indicates that the excess is established at
the $6.1 \sigma$ level of significance.  The strong signal on small velocity 
and impact parameter scales demonstrates that \civ\ absorption systems are
associated with galaxies.  Further, the lack of signal on larger velocity and
impact parameter scales strongly suggests that \civ-absorbing gas generally
arises in individual galaxies, rather than in a diffuse medium that is loosely
associated with galaxy groups, clusters, or other large-scale structures.

\subsection{Galaxy and C IV Absorber Pair Sample}

  To identify galaxy and absorber pairs that are likely to be physically
associated with each other, we followed procedures similar to those described
in Paper I.  First, we considered a galaxy and absorber pair of velocity
separation $v$ and impact parameter separation $\rho$ to be physically
associated if $\xi_{\rm ga}(v,\rho) > 1$.  In cases where more than one galaxy
could be paired with one absorber under this criterion, we chose the galaxy of
smallest impact parameter to form the pair.  Next, we excluded galaxy and
absorber pairs within 3000 \kms\ of the background QSOs (because such galaxies
and absorbers are likely to be associated with the QSOs).  Finally, we measured
$3 \sigma$ upper limits to absorption equivalent widths of galaxies that are
not paired with corresponding absorbers, retaining only those measurements with
$3 \sigma$ upper limits satisfying  $W < 0.3$ \AA.  Following these procedures,
we identified 14 galaxy and \civ\ absorber pairs and 36 galaxies that do not
produce corresponding \civ\ absorption to within sensitive upper limits.
Redshifts of the galaxy and absorber pairs range from $z = 0.1242$ to 0.8920
with a median of ${\rm med}(z) = 0.3905$, and impact parameters of the galaxy
and absorber pairs range from $\rho = 12.4$ to $142.9 \ h^{-1}$ kpc with a
median of ${\rm med}(\rho) = 52.6 \ h^{-1}$ kpc.  The $B$-band luminosities of
the galaxies range from $L_B = 0.03$ to $2.5\ L_{B_*}$.  A complete list of the
measured properties of the galaxies and \civ\ absorbers is presented in Paper
II.

\subsection{The Extent of C IV-Absorbing Gas Around Galaxies}

  To determine the extent of \civ-absorbing gas around galaxies, we examined
the relationship between rest-frame \civ\ $\lambda 1548$ equivalent width $W$
and galaxy impact parameter $\rho$.  Figure 2 shows $W$ versus $\rho$ for the
14 galaxy and absorber pairs and the 36 galaxies that do not produce
corresponding \civ\ absorption to within sensitive upper limits.  Circles
represent early-type elliptical or S0 galaxies, triangles represent early-type
spiral galaxies, and squares represent late-type spiral galaxies.  Galaxy
morphology is determined on the basis of the integrated light ratio of the disk
and bulge surface brightness profiles, as described in Papers I and II.  Open
points with arrows indicate 3 $\sigma$ upper limits to $W$ for galaxies that do
not produce corresponding \civ\ absorption lines.

  Figure 2 exhibits three interesting points: First, 12 of 18 galaxies (67\%)
of impact parameter $\rho < 70 \ h^{-1} $ kpc are associated with
corresponding \civ\ absorption lines, while only two of 32 galaxies (6\%) of
impact parameter $\rho > 70 \ h^{-1}$ kpc are associated with corresponding
\civ\ absorption lines.  Evidently, there are abrupt boundaries between the
\civ\ absorbing and non-absorbing regions of galaxies.  Second, unlike the
situation for Mg\,II- and \lya-absorbing gas around galaxies, there is no clear
trend for rest-frame equivalent width $W$ to be progressively stronger at
smaller impact parameter $\rho$.  Third, galaxies associated with corresponding
\civ\ absorption lines span a wide range of morphological type and luminosity.
Specifically, two of the 14 absorbing galaxies are elliptical or S0 galaxies
and 12 are spiral galaxies, and three of the 14 absorbing galaxies are of 
luminosity $L_B > L_{B_*}$, seven are of $0.5 L_{B_*} < L_B < L_{B_*}$, and
four are of $L_B < 0.5 L_{B_*}$.  Apparently, galaxies of a range of
morphological type and luminosity possess extended \civ\ gaseous envelopes of
radius $R \approx 100 \ h^{-1}$ kpc, with abrupt boundaries between the \civ\
absorbing and non-absorbing regions.

\subsection{The Relationship between Properties of C IV Absorption Systems and
Properties of Galaxies}

  To determine how properties of the \civ\ absorption systems depend on
properties of the galaxies, we examined the relationship between rest-frame
\civ\ $\lambda 1548$ equivalent width $W$ and galaxy impact parameter $\rho$
scaling for various properties (luminosity, surface brightness, and redshift) 
of the galaxies.  Motivated by the apparent abrupt boundary between the \civ\ 
absorbing and non-absorbing regions and the apparent lack of a $W$ versus 
$\rho$ anti-correlation described in \S\ 2.3, we modeled the distribution of 
\civ\ clouds in the extended gaseous envelopes of galaxies by uniform spheres 
of radius that depends on some property of the galaxy.  Then the number $n$ of 
\civ-absorbing clouds intercepted along the line of sight is proportional to 
the path length through the sphere:
\begin{equation}
n =2 l_0 [R^2(x) - \rho^2]^{1/2},
\end{equation}
where $l_0$ is the number density per unit length of the clouds, $R$ is the
radius of the sphere, and $x$ is some property of the galaxy.  Because \civ\
equivalent width is correlated with the number of individual absorbing
components intercepted along the line of sight (Wolfe 1986; York \etal\ 1986;
Petitjean \& Bergeron 1994), we take $W = nk$, where $k$ is the equivalent
width of a typical absorbing component.  Combining equations (1) and (2), the
\civ\ equivalent width $W$ expected at impact parameter $\rho$ is given by
\begin{equation}
W = \left\{
\begin{array}{l@{\quad \quad}l}
2 k l_0 [R^2(x)-\rho^2]^{1/2} & \rho \apl R(x) \\
0                             & \rho \apg R(x) , \\
\end{array}
\right.
\end{equation}
as a function of some property of the galaxy $x$.

  First, we established a fiducial fit of the model described by equation (2)
to the observations for the case in which the radius of the sphere does not
depend on any property of the galaxy, i.e.\ for $R = R_*$, where $R_*$ is the
absorbing radius of the galaxy.  Following procedures described in Paper I and
substituting $W$ for the dependent variable (instead of $\log W$ as in Paper
I), we solved for $k l_0$, $R_*$, and the ``cosmic scatter'' $\sigma_c$ using
a likelihood analysis.  The results are summarized in row 1 of Table 1, which
lists $k\,l_0$, $R_*$, $\sigma_c$, and an estimate of goodness of fit 
$\chi_{\nu}^2$ that is equivalent to a reduced $\chi^2$ with upper limits 
properly taken into account.

  Next, we repeated the fitting allowing the radius $R$ to scale with galaxy
$B$-band luminosity $L_B$ as
\begin{equation}
R = R_* \left( \frac{L_B}{L_{B_*}} \right)^{\alpha},
\end{equation}
where $R_*$ is the characteristic absorbing radius of an $L_*$ galaxy.  The
results are summarized in row 2 of Table 1.  After accounting for galaxy
$B$-band luminosity, $\chi_{\nu}^2$ is significantly improved with respect to 
the fiducial fit, and 
\begin{equation}
\alpha =0.5 \pm 0.1,
\end{equation}
which indicates that the dependence between $R$ and $L_B$ is established at
the 5 $\sigma$ level of significance.  The characteristic radius is
\begin{equation}
R_* = 95.9 \pm 7.0 \ h^{-1} \ {\rm kpc}.
\end{equation}
This result applies over the $B$-band luminosities $0.03 \apl L_B \apl 2.5\
L_{B_*}$ spanned by the observations.

  Figure 3 shows $W$ versus $\rho$ for the 14 galaxy and absorber pairs and the
36 galaxies that do not produce corresponding \civ\ absorption to within
sensitive upper limits after scaling for galaxy $B$-band luminosity.  The
abrupt boundary between the \civ\ absorbing and non-absorbing regions is more
evident in Figure 3 than in Figure 2.  Specifically,  after scaling for galaxy
$B$-band luminosity, 14 of 15 (93\%) of galaxies of impact parameter $\rho <
112 \ h^{-1}$ kpc are associated with corresponding \civ\ absorption lines,
while {\em no} galaxies of impact parameter $\rho > 112 \ h^{-1}$ kpc produce
corresponding \civ\ absorption to within sensitive upper limits.  A likelihood
analysis indicates that the covering factor $\kappa$ of C IV-absorbing gas
around galaxies approaches unity at impact parameters $\rho < 112 \ h^{-1}$
kpc, with a $1 \sigma$ lower bound of $\langle \epsilon \kappa \rangle = 0.91$.
The solid curve in Figure 3 shows the best-fit uniform sphere model.  

  Next, we repeated the fitting substituting galaxy surface brightness (at the
half-light radius) $\mu_e$ and redshift $(1+z)$, respectively, for galaxy 
$B$-band luminosity.  The results are summarized in rows 3 and 4 of Table 1.  
After accounting for surface brightness or redshift, the scaling exponent
$\alpha$ is indistinguishable from zero, and there is no significant
improvement in $\chi_{\nu}^2$.  We conclude that the C IV extent of galaxies
does not depend strongly on galaxy mean surface brightness or redshift.  This
result applies over surface brightnesses $18.2 \apl \mu_e \apl 25.4$ mag
arcsec$^{-2}$ and redshifts $0.09 \apl z \apl 0.83$ spanned by the
observations.

  Results of the likelihood analysis can also be used to estimate the mean
number of \civ\ clouds intercepted along a line of sight through a galaxy.
Given our estimates of the cosmic scatter $\sigma_c$ and the typical absorption
equivalent width per absorbing component per unit length $k\,l_0$ and assuming
Poisson counting statistics, so that $\sigma_c$ is related to $k$ by
\begin{equation}
\sigma_c = k (W/k+1)^{1/2},
\end{equation}
we find based on the results presented in row 2 of Table 1 that a line of sight
through the center of an $L_*$ galaxy encounters on average 
\begin{equation}
n_0 = 8.2 \pm 3.0
\end{equation}
\civ\ absorbing clouds of equivalent width $k = 0.13$ \AA.  

\section{DISCUSSION}

  The primary results of the analysis are as follows:  (1) Galaxies of a range
of morphological type and luminosity appear to possess extended \civ\ gaseous
envelopes of radius $R \approx 100 \ h^{-1}$ kpc, with abrupt boundaries
between the C\,IV absorbing and non-absorbing regions.  (2) The extent of
\civ-absorbing gas around galaxies scales with galaxy $B$-band luminosity as $R
\propto L_B^{0.5 \pm 0.1}$ but does not depend strongly on galaxy surface
brightness, redshift, or morphological type.  And (3) the covering factor of
\civ\ clouds within $\approx 100 \ h^{-1}$ kpc of galaxies is nearly unity, but
there is a large scatter in the mean number of clouds encountered along the
line of sight.  Apparently, {\em galaxies of a wide range of luminosity and
morphological type are surrounded by chemically-enriched gas\footnote{It is,
however, not clear whether or not the extended gas contains abundant metals. 
There has been a lack of chemical abundance measurements for low-redshift QSO 
absorption systems, because such analysis requires high resolution, high 
sensitivity QSO spectroscopy in the ultraviolet spectral range.  The only two 
existing chemical abundance analyses at redshifts $z<1$ are both for absorbers
of neutral hydrogen column density $\log\ N(\hI) ({\rm cm}^{-2}) \apg 16.0$ and
both systems exhibit chemical abundance that approaches the typical solar 
values (Bergeron \etal\ 1994; Chen \& Prochaska 2000).  These measurements,
however, cannot be considered representative of the majority of QSO absorption
systems at low redshifts, because of an apparent selection bias.  Because
observations of QSO absorption systems at high redshifts show a wide range of
metallicity, we believe that QSO absorption systems at low redshifts also have
a range of metallicity.  The absorbers in our sample have been identified from 
random lines of sight.  We therefore expect that statistical properties of the
absorbers in our sample are comparable to statistical properties of all
low-redshift QSO absorption systems.  Namely, there exists a range of chemical
abundance in our absorber sample.} that extends for at least $\approx 100 \ 
h^{-1}$ kpc}.  

  By combining results of previous studies of extended Mg\,II, \civ, and \lya\
gas around galaxies (derived by analysis of QSO absorption systems), we can
establish a schematic picture of the structure of extended gas around galaxies:
A typical $L_*$ galaxy is surrounded by high column density gas of radius a few
tens kpc (e.g.\ van Gorkom 1993), which, because of its high density, remains
mostly neutral.  The gas becomes gradually ionized as the gas density decreases
with increasing galactic radius.  Low-ionization species, such as singly
ionized magnesium Mg\,II, become the dominant observational signature out to
$\approx$ 50 $h^{-1}$ kpc (e.g.\ Bergeron \& Boiss\`e 1991).  As the density
continues to decline at larger galactic radii, the gas becomes still more
highly ionized.  High-ionization species, such as triply ionized carbon \civ, 
become the dominant observational signature out to $\approx$ 100 $h^{-1}$ kpc,
at neutral hydrogen column densities $N(\hI) \approx 10^{16}$ \cmjj\ (this 
paper).  The tenuous gas continues to extend to at least $\approx$ 180 $h^{-1}$
kpc, at neutral hydrogen column densities at least as low as $N(\hI) \approx 3
\times 10^{14}$ \cmjj\ (Papers I and II).  This picture applies to galaxies of
a wide range of luminosity and morphological type. 

  Establishing the origin of the chemically enriched gas at large galactic
radii is important for understanding the processes of galaxy formation and
evolution that deposit chemically enriched material far from galaxies.  Here we
consider two possible scenarios for the origin of this gas:  (1) that the gas
is driven out of the galaxies by strong galactic winds and (2) that the gas was
produced at large galactic radii.  

  In the first scenario (which is known as the galactic fountain model), hot
gas is brought to large galactic distances by violent supernova explosions, 
forms cold clouds through subsequent radiative cooling, and falls back to the
center of the galaxy (Corbelli \& Salpeter 1988).  Comparing with the results
of our analysis, we find that the galactic fountain model is unlikely to
account for extended \civ\ gas around galaxies for the following reasons:
First, hot gas driven by supernova explosion can reach to at most $\approx$ 20 
kpc (e.g.\ Corbelli \& Salpeter 1988) before it completely escapes the galaxy, 
but our analysis indicates that \civ-absorbing gas extends to $\approx$ 100
$h^{-1}$ kpc.  Second, the supernova explosion rate varies significantly
between galaxies of different morphological types, but our analysis suggests
that galaxies of all morphological types possess extended \civ-absorbing gas to
large radii.  Third, given a fixed initial flow speed, the ejected gas may
reach larger galactic radii from within smaller galaxies, which is at odds with
the scaling relation of our analysis that the \civ\ gaseous extent increases
with galaxy luminosity.  

  In the second scenario, chemically enriched gas is deposited at large radii
either by the first generation stars (Population III) or through progressive
accretion of gas from surrounding satellite galaxies.  But because the
pregalactic enrichment due to Population III stars is strongly constrained by
the metallicity of Population II stars to be less than $Z\approx 10^{-3}$ 
(Carr, Bond, \& Arnett 1984) and because observations show that there is 
already a wide range of metallicity in high-redshift QSO absorption systems 
(e.g.\ Rauch 1998), we conclude that Population III stars cannot explain all of 
the \civ\ absorbing clouds at large radii, although they may contribute to some 
of them.  On the other hand, theoretical analysis of an accreting satellite 
model has shown that chemically enriched gas may be stripped out of the 
surrounding satellites through tidal interaction or may be ejected from the 
satellites due to supernova explosions to form absorbing clouds at large radii 
of the primary galaxy (Wang 1993).  Supporting evidence is found from 
observations of nearby spiral galaxies, which shows that there exists at least 
one satellite around each Milky Way type galaxy (Zaritsky \etal\ 1993).  
In addition, \hI\ 21 cm analysis of local dwarf spheroidals suggests that the 
presence of satellite galaxies may contribute to extended gas in galactic halos 
(Blitz \& Robishaw 2000).

  We believe that it is therefore likely that accreting satellites may account
for chemically enriched gas at large galactic distances, which has implications
for the dynamics of absorbing gas clouds as follows:

  First, for an isothermal sphere, the density of accreted gas decreases with
radius $R^{-2}$.  But because of dissipational collisions, absorbing clouds may
gradually lose angular momentum during the cooling process and spiral down to
the primary galaxies, causing the density of accreted gas to fall off more
rapidly than $R^{-2}$ (Silk \& Norman 1981).  It appears that the abrupt
boundary between the C\,IV absorbing and non-absorbing regions apparent in
Figure 3 supports a sharp steepening in the number density of \civ-absorbing
clouds in the outer gaseous envelopes of galaxies.  Fitting $W$ versus $\rho$
of Figure 3 with a single power-law model $W \propto \rho^{\alpha}$ yields
$\alpha=-1.45 \pm 0.18$, which implies a density distribution $R^{-2.45 \pm
0.18}$ (see Lanzetta \& Bowen 1990).  The dotted line in Figure 3 shows the
best-fit power-law model.  We conclude that the abrupt boundaries the C\,IV
absorbing and non-absorbing regions of galaxies is indicative of dissipational
formation processes of galaxies.

  Second, our results indicate that the extent of \civ-absorbing gas does not
depend strongly on galaxy redshift, which implies that \civ\ gaseous envelopes
of galaxies are essentially static over a large cosmic time interval.  Given
our estimate in Equation (7) of the average number of clouds along a line of
sight through the center of an $L_*$ galaxy and adopting a velocity dispersion
$v \approx 200$ \kms\ of a typical $L_*$ galaxy, we estimate that the collision
time between \civ-absorbing clouds is $t_{\rm coll}\approx 2 R_* / (n_0 v)
\approx 1.1 \times 10^8$ yr.  In comparison, the dynamical time for the clouds
to fall through the gravitational potential of a galaxy is $t_{\rm dyn} \approx
\pi R_*/2 v \approx 7.4 \times 10^8$ yr.  Unless the gaseous envelopes of
galaxies are continually supplied with matter or energy, it is clear that the
accreted gas clouds will rapidly decay to lower orbits, resulting in smaller
gaseous cross sections with time.  This directly contradicts our observations.
A solution to this problem may be that satellites at large galactic radii
provide a constant supply of fresh gas.  We therefore conclude that if
\civ-absorbing clouds are formed through dissipational accretion of gas from
surrounding satellites, the mean separation between the satellites and the
primary galaxies may be larger than 100 $h^{-1}$ kpc.

\acknowledgements

  This work was supported by STScI grant GO-07290.01-96A and NSF grant 
AST-9624216.

\newpage

\figcaption{Galaxy--\civ\ absorber cross-correlation function $\xi_{ga}(v,
\rho)$ vs. velocity separation $v$ and impact parameter separation $\rho$.  
The bin size in velocity separation is 250 \kms.  The bin size in impact
parameter separation is 100 $h^{-1}$ kpc.  Error bars indicate 1 $\sigma$ 
Poisson counting errors.}

\figcaption{Logarithm of \civ\ rest-frame equivalent width $W$ vs.\ logarithm 
of galaxy impact parameter $\rho$.  Circles represent early-type elliptical or 
S0 galaxies, triangles represent early-type spiral galaxies, and squares 
represent late-type spiral galaxies.  Closed points indicate detections.  Open 
points with arrows indicate $3 \sigma$ upper limits of non detections.}

\figcaption{Logarithm of \civ\ rest-frame equivalent width $W$ vs.\ logarithm 
of galaxy impact parameter $\rho$ scaled by galaxy $B$-band luminosity. The 
scaling factor is determined from the analysis described in \S\ 3.4.  Symbols 
represent the same as those in Figure 2.  The solid curve indicates a uniform 
sphere model that best fits the data.  The dotted line indicates the best-fit 
power-law model to the data.}

\newpage 

\plotone{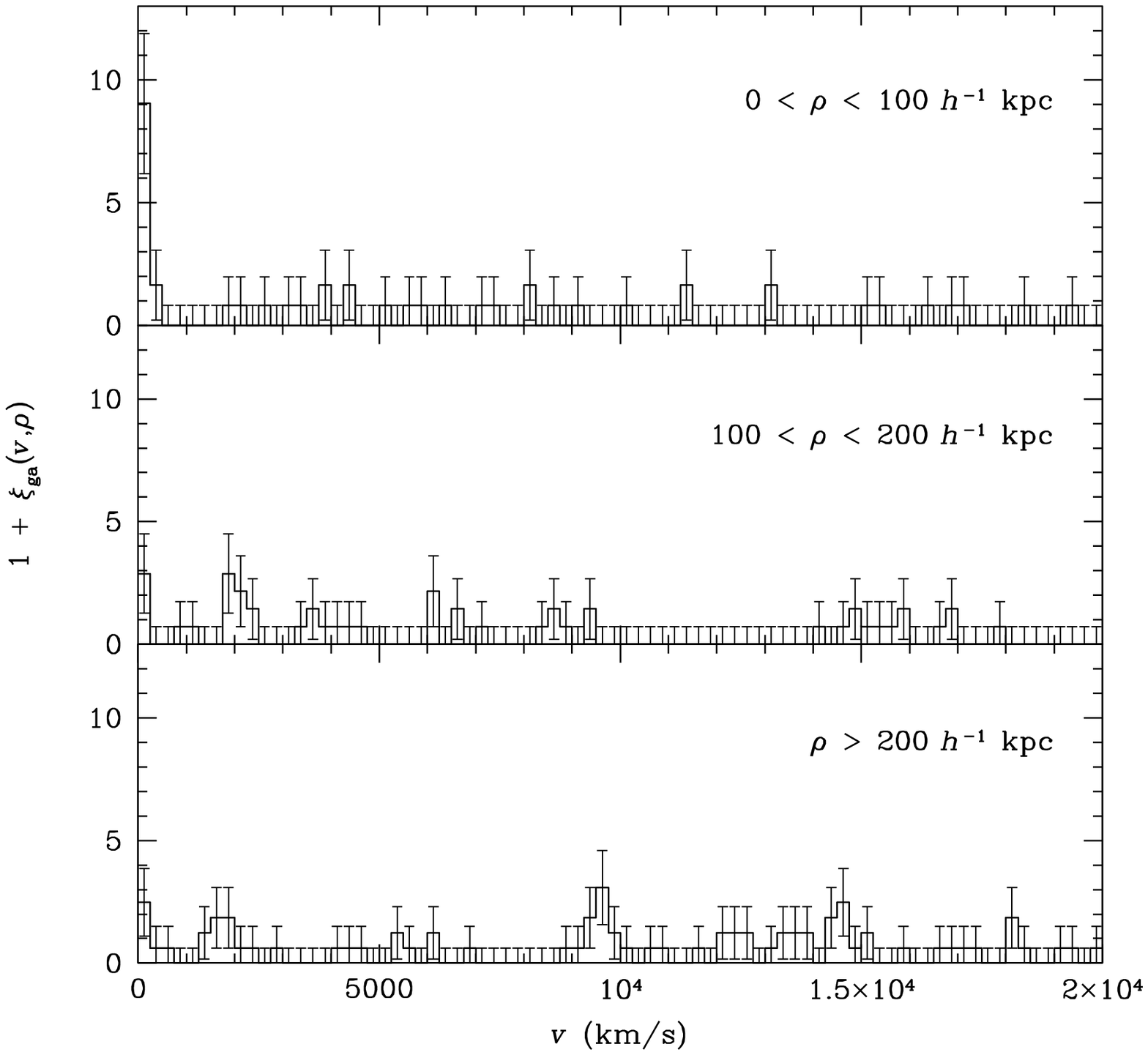}

\newpage 

\plotone{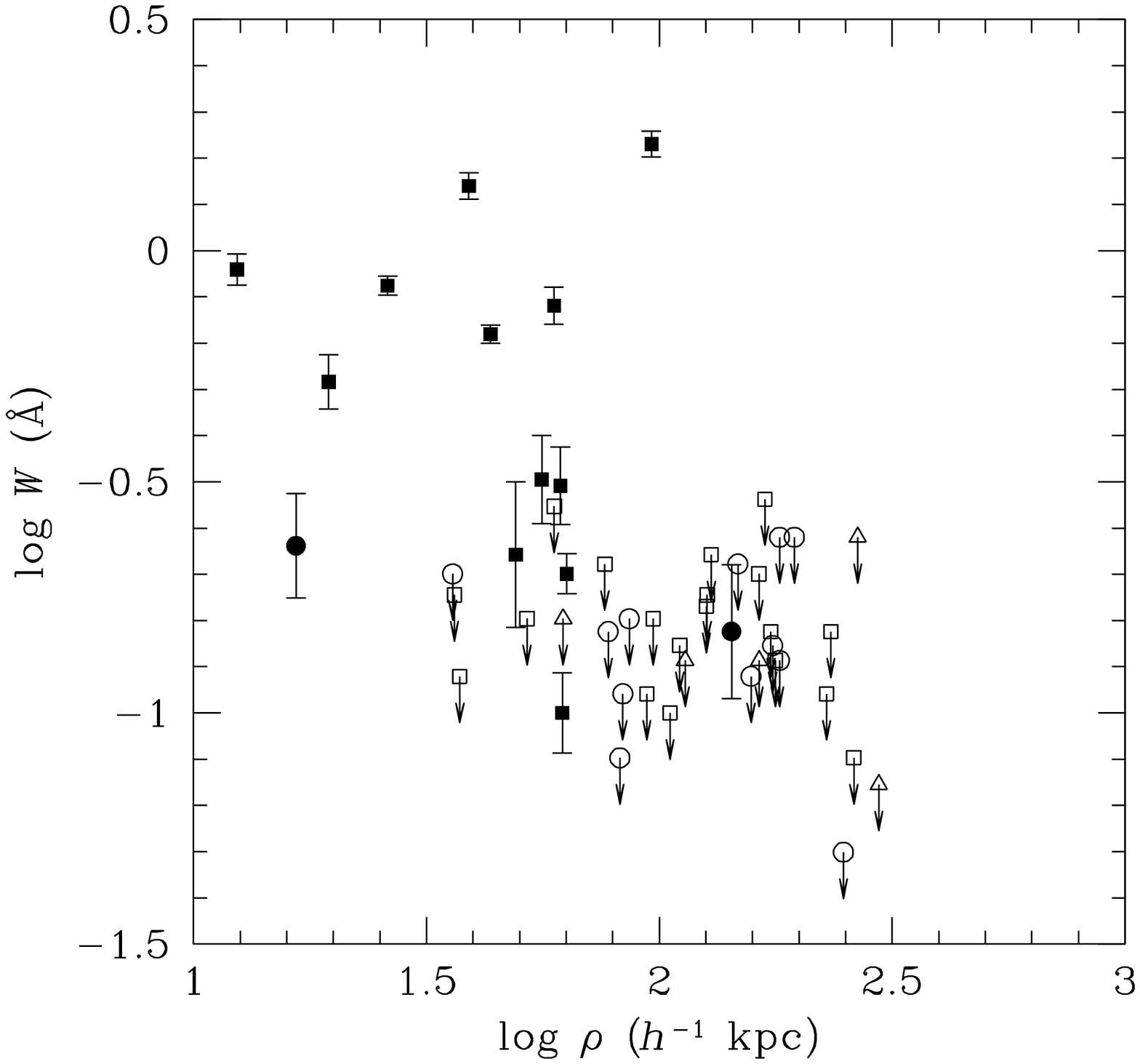}

\newpage 

\plotone{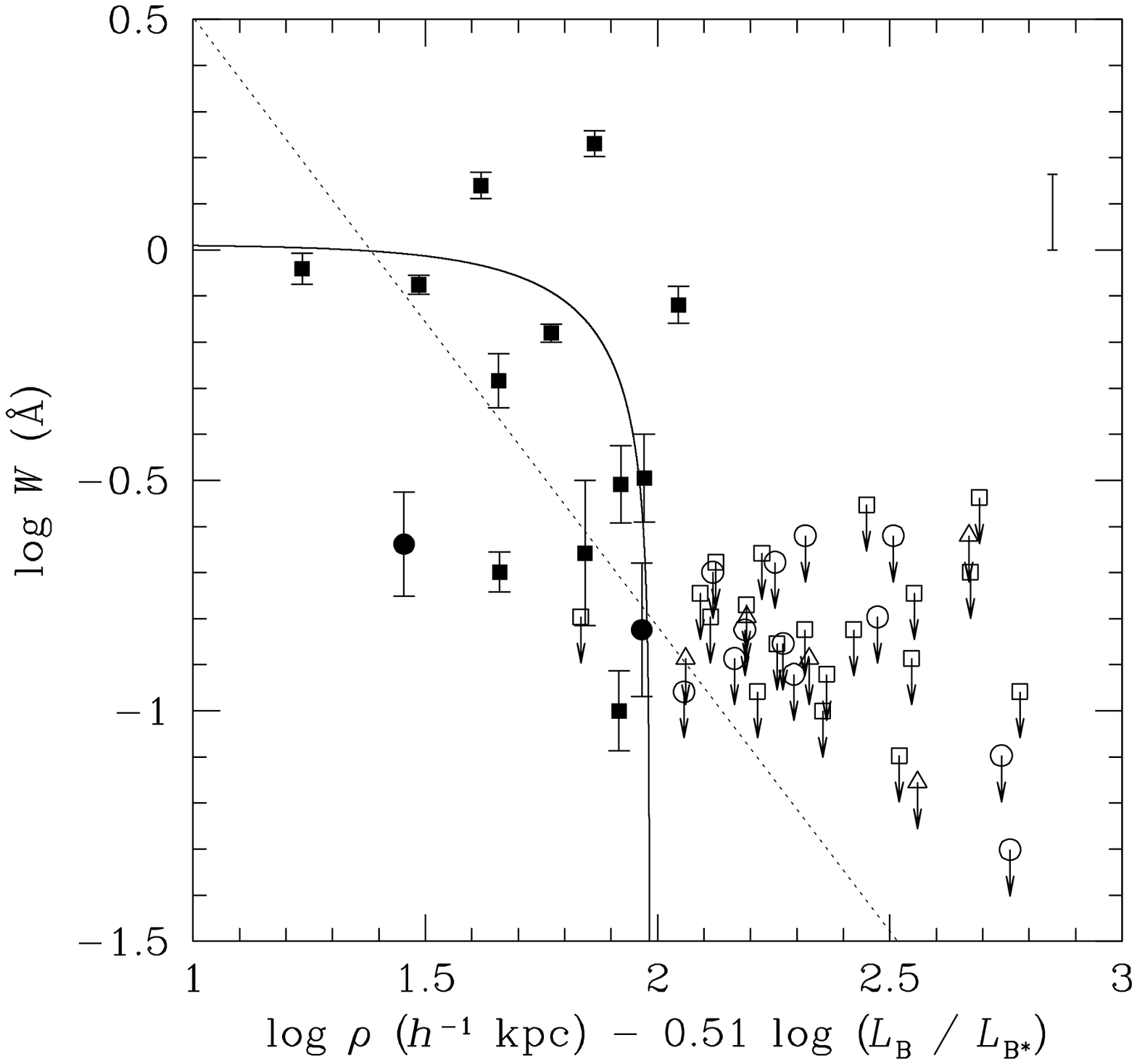}



\newpage 

\begin{table}
\begin{center}
\begin{tabular}{p{2.00 in}clccc}
\multicolumn{6}{c}{Table 1} \\
\multicolumn{6}{c}{Results of Likelihood Analysis} \\
\hline
\hline
Measurements & $k\,l_0$ & \multicolumn{1}{c}{$R_*$} & $\alpha$ & $\sigma_c$ & 
$\chi_{\nu}^2$ \\
\hline
1. $W - \rho$ \dotfill & $0.005 \pm 0.001$ & $64.0 \pm 2.4$ 
	& ... & 0.23 & 2.9 \\
2. $W - \rho - L_B$ \dotfill & $0.006 \pm 0.001$ & $95.9 \pm 7.0$ 
	& $0.5\pm 0.1$ & 0.39 & 1.2 \\
3. $W - \rho - \mu_e$ \dotfill & $0.004 \pm 0.001$ 
        & $66.2 \pm 4.0$ & $0.04 \pm 0.05$ & 0.25 & 2.7 \\
4. $W - \rho - (1+z)$ \dotfill & $0.005 \pm 0.002$ & $60.0 \pm 3.0$
        & $0.4 \pm 0.7$ & 0.29 & 2.7 \\
\hline
\end{tabular}
\end{center}
\end{table}

\end{document}